\documentclass[floatfix,a4paper,prl,twocolumn,superscriptaddress,amsmath,amssymb,footinbib]{revtex4-1}
\pdfoutput=1

\usepackage{natbib}
\usepackage[english]{babel}
\usepackage{letltxmacro}
\LetLtxMacro{\ORIGselectlanguage}{\selectlanguage}
\makeatletter
\DeclareRobustCommand{\selectlanguage}[1]{%
  \@ifundefined{alias@\string#1}
    {\ORIGselectlanguage{#1}}
    {\begingroup\edef\x{\endgroup
       \noexpand\ORIGselectlanguage{\@nameuse{alias@#1}}}\x}%
}
\newcommand{\definelanguagealias}[2]{%
  \@namedef{alias@#1}{#2}%
}
\makeatother
\definelanguagealias{en}{english}
\definelanguagealias{English}{english}

\usepackage{layouts} 
\usepackage{graphicx}
\usepackage{soul}
\usepackage[usenames,dvipsnames]{color}

\usepackage[colorlinks=true,allcolors=blue]{hyperref}

\usepackage{bbm}

\newcommand{\mean}[1]{\left\langle #1 \right\rangle}	

\newcommand{\abs}[1]{\ensuremath{\left\vert #1 \right \vert}} 
\newcommand{\bra}[1]{\langle #1 |}
\newcommand{\ket}[1]{| #1 \rangle}
\newcommand{\braket}[2]{\langle #1|#2 \rangle}

\newcommand{\tr}[2][]{\ensuremath{\text{tr}_{#1}\left[ #2\right]}}

\newcommand{\tone}{\mathbbm{1}}

\newcommand{\refeq}[1]{Eq.~(\ref{#1})}

\def\clap#1{\hbox to 0pt{\hss#1\hss}}

\begin{document}

\title{Construction and universal application of entanglement erasing partner states}

\author{Daniel Hetterich}
\affiliation{Institute for Theoretical Physics,  University of W\"urzburg, 97074 W\"urzburg, Germany}
\affiliation{Department of Physics, Technical University of M\"unchen, 85748 Garching, Germany}

\author{Polina Matveeva}
\affiliation{Department of Physics and Research Center Optimas, Technical University of Kaiserslautern, 67663 Kaiserslautern, Germany}

\date{\today}

\begin{abstract}
We investigate the subadditivity of the bipartite entanglement entropy (EE) of many-particle states, represented by Slater determinants, with respect to single particle excitations. 
In this setting, subadditivity can be phrased as erasure of EE, i.e. as a relative decrease in EE when adding excitations to the quantum state. 
We identify sets of single particle states that yield zero EE if jointly excited. Such states we dub entanglement erasing partner states (EEPS). 
These EEPS reveal a mechanism that describes how  to disentangle two subspaces of a Hilbert space by exciting additional states. 
We demonstrate this general finding in Anderson and many-body localized models. 
The studied concept of entanglement erasure further enables us to derive the EE of Slater determinants in the free tight binding model. Here, our analytical findings show surprisingly good agreement with numerical results of the interacting XXX chain. 
The described EEPS further impose a universal, i.e. model independent, erasure of EE for randomly excited Slater determinants. This feature allows to compute many-particle EE by means of the associated single particle states and the filling ratio. This novel finding can be employed to drastically reduce the computational effort in free models.
\end{abstract}

\maketitle

The purely quantum phenomenon of entanglement has impacted the whole history of quantum mechanics and is to date an active topic of research. 
While entanglement first demonstrated the non-local nature of quantum physics~\cite{Einstein1935}, Bell's inequalities later excluded a description of quantum mechanics by means of classical hidden variables~\cite{Bell1964}. In addition, Bell's work first quantified the quantum correlations between two subsystems that arise due their entanglement.

Today quantum entanglement plays a key role in research fields ranging from  black holes~\cite{Bombelli1986,Srednicki1993}, photosynthetic processes~\cite{Sarovar2010}, and is a central concept in quantum information theory~\cite{Nielsen2011}. 
Entanglement further is extremely valuable for characterizing ground- and excited states of many-body systems and their behavior after quenches~\cite{Lieb1972,Shi2003,Amico2008,Alba2009,Eisert2010,Cheneau2012,Ganahl2012, Ferenc2012,Collura2013,Paula2017}. For instance, the logarithmic growth of entanglement after a quench \cite{Znidaric2008,Bardarson2012,Serbyn2014,
Hetterich2017,DeTomasi2017,Smith2017} is considered as a defining feature of the many-body localized phase. 
The entanglement of many-body states also shows intriguing features that depend on the number of (quasi) particles, both, in interacting ~\cite{Berkovits2013, Molter2014} and free models~\cite{Berkovits2013, Storms2014, Ramarez2014} and it is an ongoing process to extend its understanding by means of quasi-particle pictures~\cite{Pizorn2012,CastroAlvaredo2018}. 

A common measure of entanglement bases on the von Neumann entropy
$\mathcal{S}(\rho) = - \tr{\rho \log_2 \rho}$
of a density matrix~$\rho$. If the considered Hilbert space $\mathcal{H}$ is divided into two parts $A,B$, i.e. $\mathcal{H} = \mathcal{H}_A \otimes \mathcal{H}_B$, for instance referring to the left and right half of a system, Araki and Lieb~\cite{araki1970} have proven the subadditivity $\mathcal{S}(\rho) \leq \mathcal{S}(\rho_A) + \mathcal{S}(\rho_B)$, where $\rho_A = \tr[B]{\rho}$ and $\rho_B=\tr[A]{\rho}$ describe the quantum system in the subspaces $\mathcal{H}_A$ and $\mathcal{H}_A$, respectively. This feature indicates entanglement between these two subsystems: The joint system has less entropy than its individual parts, hence, these parts must be correlated. This quantum correlation is quantified by the entanglement entropy (EE), for instance for a pure state $\ket{\psi}\in\mathcal{H}$ given by 
\begin{equation} \label{eq:von_neumann}
S_E(\ket{\psi}) = -\tr{\rho_A \log_2 \rho_A},
\end{equation}
with  $\rho_A = \tr[B]{\,\ket{\psi}\bra\psi}$. 

In this work, we explore how this EE depends on the particle number of the state $\ket{\psi}$. Specifically, we study how the EE of many-particle states described by Slater determinants behaves when an additional fermionic excitation is added. We find in general that EE behaves subadditive in such a process, i.e. the EE of a Slater determinant is lower than the bare sum of the EEs of the used single particle states. We interpret this subadditivity as a mutual \emph{erasure} of EE between these single particle states. 
By means of quantifying this erasure of EE we identify quantum states that yield zero EE if they are excited simultaneously. Such sets of states we dub entanglement erasing partner states (EEPS). Using EEPS we demonstrate how to uncorrelate two parts of a Hilbert space by exciting additional states. For instance, in Anderson (or many-body) localized models we find that EEPS are built from only a few particles, which makes the scheme of uncorrelating by exciting particles especially effective here.  For the non-interacting tight binding model our analysis enables to derive concrete values of Slater determinants of single particles. For such states we find that the possible values of EE are discrete, where the minimum possible EE of two particles is larger than zero. Our derived discrete values of EE explain previous numerical observations made in the interacting XXX spin 1/2 chain~\cite{Molter2014}.  
Moreover we find a model independent relative erasure of EE for $N$ randomly excited single particle states. This remarkable effect can be fully understood by means of our described EEPS.

\emph{Erasure of EE ---}
Before we discuss these results in detail, we quantify the erasure of EE between single particle states and study its dependency on these states. 
To this end, just two orthogonal fermionic states $\ket{1}=c_1^\dagger\ket{\emptyset}$ and $\ket{2}=c_2^\dagger\ket{\emptyset}$ are needed. Here, $\ket{\emptyset}$ describes the empty system of zero particles. This defines the state $\ket{1,2}=c_1^\dagger c_2^\dagger \ket{\emptyset}$, which describes two identical particles, each being in one of the previous defined states. For an arbitrary bipartition $\mathcal{H}=\mathcal{H}_A\otimes \mathcal{H}_B$, the EE then yields
\begin{align}\label{eq:subadditivity}
&S_E(\ket{1,2}) \leq S_E(\ket{1}) + S_E(\ket{2}).
\end{align} 
To see this, the relevant step is to express the single particle states by 
\begin{align} \label{eq:schmidt-decoposition}
\ket{i} &= \sqrt{\lambda_i} \ket{i}_A \otimes \ket{\emptyset}_B + \sqrt{1-\lambda_i} \ket{\emptyset}_A \otimes \ket{i}_B\\
&=\sqrt{\lambda_i} \ket{i}_A + \sqrt{1-\lambda_i} \ket{i}_B,
\end{align}
where $\ket{i}_{A}, \ket{\emptyset}_A \in \mathcal{H}_A$ and  $\ket{i}_{B},\ket{\emptyset}_B\in \mathcal{H}_B$ are single and zero particle wavefunctions on $\mathcal{H}_A$ and $\mathcal{H}_B$, respectively. In the second line, we introduce a handy notation that simplifies further expressions. By employing this construction, the positive value $\lambda_i$ equals the probability that the particle of state $\ket{i}$ is found in the subspace $\mathcal{H}_A$. Also note that Eq.~\eqref{eq:schmidt-decoposition} is a Schmidt decomposition of $\ket{i}$ with the two Schmidt coefficients $\sqrt{\lambda_i}$ and $\sqrt{1-\lambda_i}$. Its single particle entanglement entropy is given by~\cite{Peschel2009} 
\begin{align}\label{eq:singleparticleentropy}
S_E(\ket{i}) &= s(\lambda_i) \\
\text{with}\; s(x) &:= -x \log_2 x - (1-x) \log_2 (1-x).
\end{align}
This determines the right-hand side of Eq.~\eqref{eq:subadditivity}. In turn, the EE of $\ket{1,2}$ is given by  
 \begin{equation}
 S_E(\ket{1,2}) = s(\nu_1) + s(\nu_2)
 \end{equation}
 with the modified squares of  Schmidt coefficients
 \begin{align}
\nu_{1,2} &= \lambda_{1,2} \pm \frac{1}{2}\left[ \sqrt{(\lambda_1-\lambda_2)^2+ 4 \lambda_1\lambda_2 \sigma} - (\lambda_1-\lambda_2)\right] \nonumber\\
 &\text{and}\; \sigma = \abs{ \bra{1}_A \cdot \ket{2}_A}^2,
\end{align} 
which we prove in the supplementary material~\footnote[0]{See supplementary material for the derivation of the proof, the derivation of the erasure, and details on the numerical procedures.}.
Note that the overlap $\sigma$ is determined by the parts of the two single particle states $\ket{1}$ and $\ket{2}$ \emph{within} subspace $\mathcal{H}_A$, which can be finite even though the states are orthogonal to each other $\braket{1}{2}=0$ in $\mathcal{H}=\mathcal{H}_A\otimes\mathcal{H}_B$.
One directly sees that for $\sigma=0$, the Schmidt coefficients remain unchanged and Eq.~\eqref{eq:subadditivity} turns into an exact equality, i.e., the EE is additive.  Similar dependencies on orthogonality in subspaces have been observed before where superpositions of states were studied~\cite{Linden2006}.

On the other hand, for $\sigma = 1$ where the two orthogonal states $\ket{1},\ket{2}$ also equal each other \emph{within} $\mathcal{H}_A$, the Schmidt coefficients become $\sqrt{\nu_1} = \sqrt{\lambda_1+\lambda_2}$ and $\sqrt{\nu_2} = 0$. We show~\cite{Note0} that this yields the minimal value of $S_E(\ket{1,2})$ for given $\lambda_1,\lambda_2$. Therefore, $\sigma=1$ implies maximal erasure of EE. If further both probabilities are connected via $\lambda_j = 1-\lambda_i$, EE is entirely erased, $S_E(\ket{i,j})=0$. This means that the quantum correlations between two halves of a system, originating from a particle $\ket{1}$, can be completely annihilated by exciting a second fermionic particle $\ket{2}$. This observation is important as it allows us below to introduce the notion of entanglement erasing partner states, which in turn is crucial for explaining the universal properties of EE of randomly excited states.

As an illustrative example of full EE erasure, let  $\ket{a}_A\in \mathcal{H}_A$ and $\ket{b}_B\in\mathcal{H}_B$ be two fermionic states. The two orthonormal states $\ket{1} = \sqrt{0.3}\ket{a}_A - \sqrt{0.7}\ket{b}_B$ and $\ket{2} = \sqrt{0.7}\ket{a}_A + \sqrt{0.3}\ket{b}_B$ both yield $S_E(\ket{1})=S_E(\ket{2})\approx 0.88\,\text{bit}$ (by writing 'bit', we emphasize the usage of the logarithm to the base 2). However, the two particle state $\ket{1,2} = \ket{a}_A\otimes \ket{b}_B$ is a product state with respect to $\mathcal{H}_A$ and $\mathcal{H}_B$, implying zero EE. 
We show next that such strong erasure of EE is also observable in important models of condensed matter physics.

\begin{figure}
\centering
\includegraphics[width=0.98\linewidth]{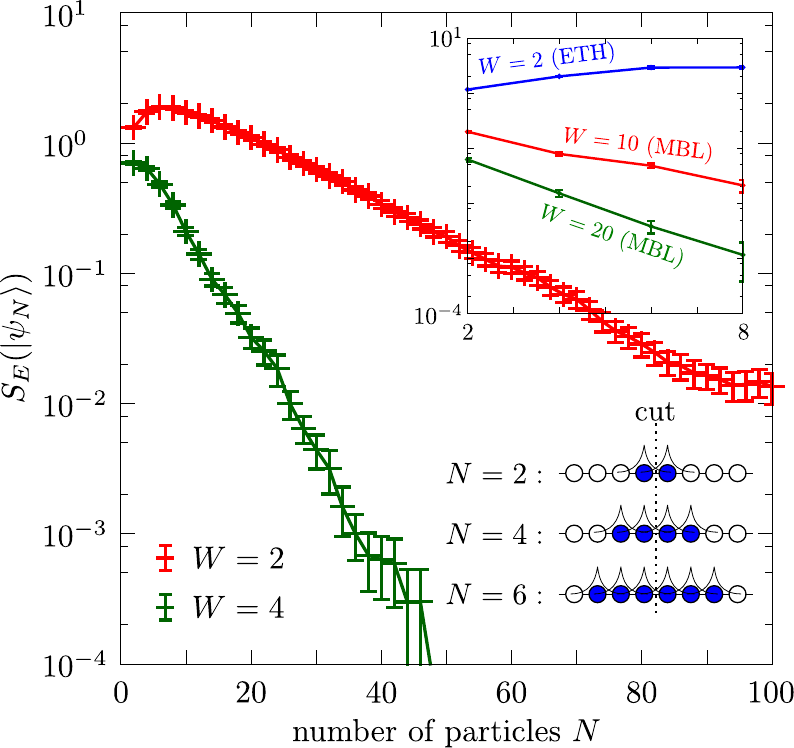}
\caption{Exponential erasure of EE by exciting additional states. Out of all $L=4096$ Anderson localized single particle states, we excite those $N$ states that are localized closest to the cut that defines the bipartite EE (see scheme in the bottom right corner). The data shows that the erasure of entanglement can dominate over the gain of entanglement if additional states are excited. In this model, the disorder strength $W$ determines the localization length and therefore the number of states within a set of EEPS. The upper inset shows the corresponding interacting model with $L=16$. There, erasure of EE can be used to decouple the bipartitions within the MBL phase only.}
\label{fig:erasure_anderson}
\end{figure}

\emph{EEPS in localized models---} Free fermions constraint to a one-dimensional chain of sites $i$ with random potentials $h_i\in[-W,W]$, described by the Hamiltonian 
\begin{equation}
H_\text{An} = - t\sum_{i=1}^L  \left(c_i^\dagger c_{i+1}+\text{h.c.} \right) + \sum_{i=1}^L h_i c_i^\dagger c_i,\label{eq:anderson}
\end{equation}
are known to experience Anderson localization~\cite{Anderson1958} for any finite disorder strength $W>0$ in the thermodynamic limit $L\to\infty$~\cite{Anderson1979, Mott1961}. Then, all single particle eigenstates $\ket{e_i^\text{An}}$ are exponentially localized around site a $i$. As a consequence, the only states that yield a finite EE between two contiguous halves of the system are those that are localized near the cut between these two halves. 
In order to demonstrate the effect of single particle subadditivity described by Eq.~\eqref{eq:subadditivity}, we therefore excite $N$ eigenstates $\ket{e_i}$ that are as close to the cut as possible.  This procedure also maximizes their mutual overlaps $\sigma$ within a half-chain bipartition and thus erases most EE. 
This yields a joint  EE that decays with particle number as $S(\ket{\psi_N}) \sim e^{-\lambda N}$, where $\lambda$ depends on the disorder $W$, see Fig.~\ref{fig:erasure_anderson}. Remarkably, the two halves of a chain can thus be exponentially fast disentangled by simply exciting additional states, i.e. adding particles into eigenstates near the cut. 

This result is not limited to the free model: Adding the density-density interaction  $V=2t \sum_i c_i^\dagger c_i c_{i+1}^\dagger c_{i+1}$  to \refeq{eq:anderson}, where $t$ is the hopping constant,  a many-body localization transition occurs around $W_c\approx 3.5\cdot 2t$ in the thermodynamic limit~\cite{Basko2006a,Oganesyan2007,Pal2010,Luitz2015}. Note that the full Hamiltonian still conserves the number of particles, which still allows us to study EEs $S_E(\ket{\psi_N})$ of $N$-particle states $\ket{\psi_N}$. Analogously to the free model, we excite those many-particle eigenstates $\ket{\psi_N}$ where the particles are localized as close to the cut as possible. Again we find exponential erasure of EE in the localized phase $W>W_c$, see lower inset of Fig.~\ref{fig:erasure_anderson}. Instead, for $W<W_c$ where the eigenstates of the interacting model are extended, erasure is weaker and not sufficient to uncorrelate the subspaces.  We continue by quantifying the erasure of EE in such extended models.

\begin{figure}
\centering
\includegraphics[width=0.98\linewidth]{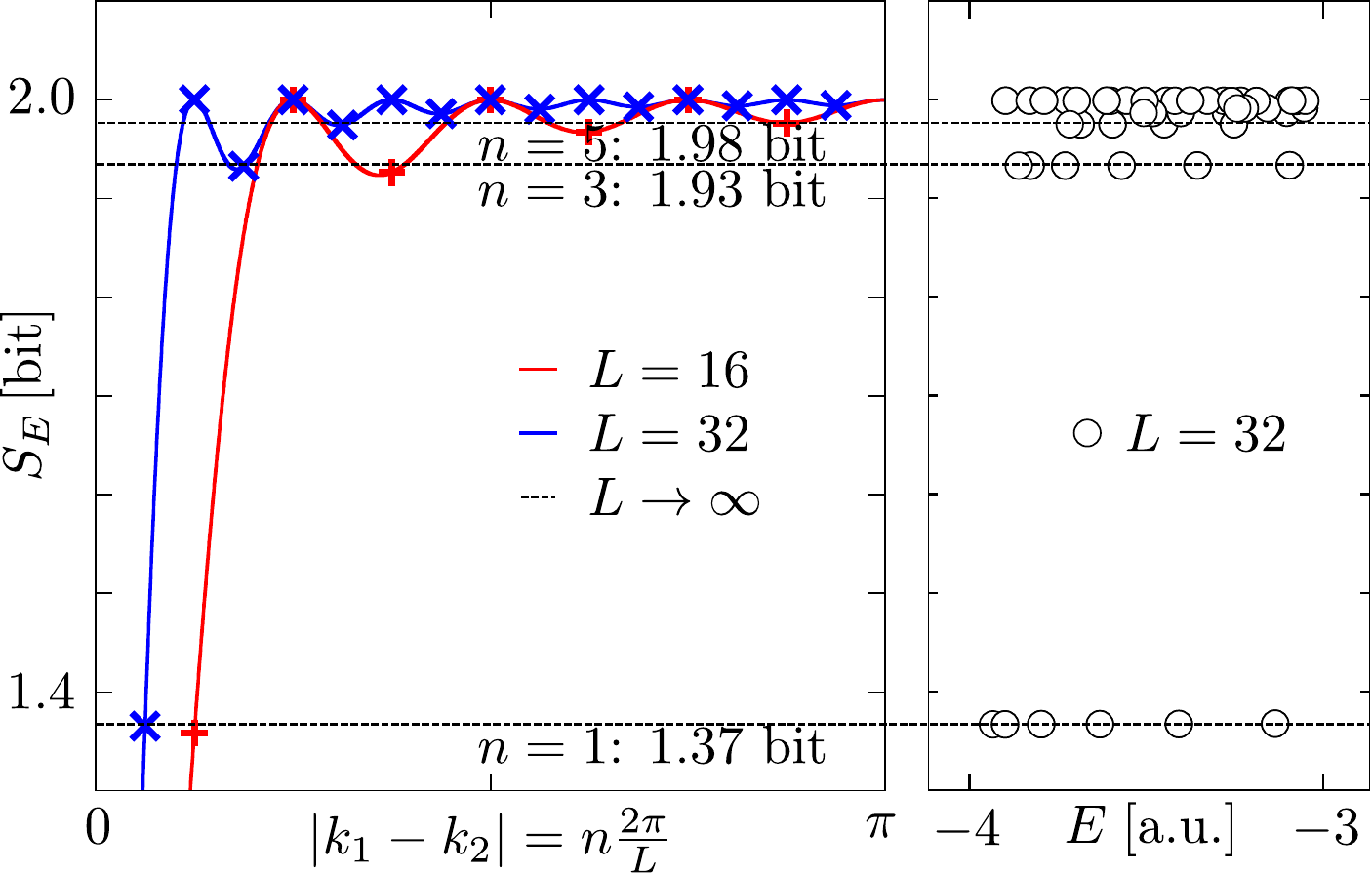}
\caption{left: Joint half-chain EE $S_E$ of two eigenstates $\ket{k_1}, \ket{k_2}$ of the free tight-binding model, according to Eqs.~\eqref{eq:entanglementtb}-\eqref{eq:entanglementtc}. Data points in respect the quantization of momentum, while the curves show the dependency of $S_E$ on $\abs{k_1-k_2}$ without this restriction. 
In the free model, the erasure of $S_E$ depends on $n\in\mathbb{N}$, where $k_1-k_2= 2n \pi/L$. For $L\to\infty$, we derive the possible values of EE that are in good agreement with  numerical data of the more complicated XXX chain~\cite{Molter2014} shown on the right side. \\
right: Half-chain EE of low-energetic two-particle eigenstates of the interacting isotropic spin $1/2$ chain, taken from~\cite{Molter2014}. }
\label{fig:tightbinding}
\end{figure}

\emph{Exact EE in the tight-binding model ---} We now employ our method in order to derive the EE of Slater determinants in the tight-binding model, described by Eq.~\eqref{eq:anderson} with no disorder, i.e. $h_i = 0$ for all $i$. As we show and discuss, our results also well-describe the many-particle eigenstates of the interacting XXX chain. Firstly, the single particle  eigenstates of the free model are
\begin{equation}
\ket{k} = \frac{1}{\sqrt{L}} \sum_{j=1}^L e^{i k j} c_j^\dagger \ket{\emptyset}
\end{equation} 
for periodic boundary conditions. 
Again let $\mathcal{H}_A$ describe one half of the chain in real space. Two states  $\ket{k_1}, \ket{k_2}$ separated by momentum $k_2 - k_1 = n 2\pi/L$, then yield~\cite{Note0}
\begin{align}\label{eq:entanglementtb}
 S_E^\text{tb}(n) &=
 s\left(1/2 +\sigma/2
\right) + s\left(1/2 - 
\sigma/2
\right), \;\text{with}\\ \label{eq:entanglementtc}
 \sigma &= \frac{2 \sin (n\pi  / 2)}{n \pi}
\end{align}
for $L\to\infty$. Note how  
the overlap $\sigma$ of $\ket{k_1}$ and $\ket{k_2}$ within $\mathcal{H}_A$ vanishes for all even values of $n$, implying an additive EE, i.e. $S_E^\text{tb}(n)=2\,\text{bit}$. For odd values of $n$, erasure occurs and the EE decreases as $n^{-2}$ for $n\gg 1$, 
\begin{equation} \label{eq:tbentropy}
S_E^\text{tb}(n)= 2 - \left(\frac{2}{n \pi \sqrt{\log 2}}\right)^2+\mathcal{O}\left(\frac{1}{n^4}\right),
\end{equation}
such that for most pairs of eigenstates $\ket{k_1},\ket{k_2}$ the EE shows an almost additive behavior. 
Most erasure is present if the momenta $k_1$ and $k_2$ are similar. Albeit the two momenta can be chosen  arbitrarily close to each other for $L\to\infty$, we find that  EE entropy can never be erased entirely,  but here we find a minimum of 
$S_E^\text{tb}(n=1) \approx 1.3675\,\text{bit}$. 
The odd integers $n$ yield  all possible \emph{discrete} values of the EE of two particle eigenstates. These values are illustrated by dashed lines in Fig.~\ref{fig:tightbinding}. For comparison, we  evaluate Eq.~\eqref{eq:entanglementtb} for continuous values of $n$ in  the left half of Fig.~\ref{fig:tightbinding}. 
 The described discrete bands of the two-particle EE have also been observed in the XXX chain, which  is the interacting analogon of our studied tight-binding model~\cite{Molter2014}. As the right half of Fig.~\ref{fig:tightbinding} shows, the low energy eigenstates of this interacting model yield very similar values of EE. This is because the low-energy spectrum of the XXX chain may be well-described by an effective free particle model, which can for example be seen by means of bosonization techniques~\cite{Gogolin}. 

\emph{Universal erasure---} 
As we have illustrated, the amount of EE erasure strongly depends on the structure of the excited (eigen)states. Yet, we next show a model-independent behavior of the \emph{relative} erasure of EE, making use of the fact that oftentimes one is interested in the EE of randomly drawn eigenstates. 
 Before we generalize, we illustrate the underlying mechanism by slightly expanding a previously discussed example. Let  $\{\ket{a_i}_A\}$ and $\{\ket{b_i}_B\}$ be orthonormal bases of the single particle spaces $\mathcal{H}_A$ and $\mathcal{H}_B$. Now define the 
 Bell states
\begin{align}
\ket{e_{2i}^\text{Bell}} &= \frac{1}{\sqrt{2}} \left( \ket{a_i}_A + \ket{b_i}_B\right)\\
\ket{e_{2i+1}^\text{Bell}} &= \frac{1}{\sqrt{2}} \left( \ket{a_i}_A - \ket{b_i}_B\right),
\end{align}
which individually show maximum EE (1~bit). Also note that two states $\ket{e_{2i}^\text{Bell}}$ and $\ket{e_{2i+1}^\text{Bell}}$ with same index $i$ form a EEPS set, i.e. they erase each others EE completely. Importantly, states with different index $i$ cannot interfere with each other, as they are also orthogonal within the subspaces $\mathcal{H}_A$ and $\mathcal{H}_B$. We now again excite $N$ single particle states by means of a Slater determinant, i.e. $\ket{\phi_N}=e_{j_1}^\dagger e_{j_2}^\dagger\ldots e_{j_N}^\dagger\ket{\emptyset}$, but this time at random. We then in general quantify the \emph{relative} amount of erasure of EE by means of the erasure factor 
\begin{equation}\label{eq:erasurefactor}
r_\infty := \frac{\mean{\,S_E(\ket{\phi_N})\,}}{\mean{\sum\limits_i^N S_E(\ket{e_i})}},
\end{equation}
where  $\mean{\;}$ denotes the expectation value of a random variable. Further, $r_\infty$
 is to be evaluated in the thermodynamic limit $L\to\infty$ and for fixed filling ratio $N/L$ . For the above defined Bell states, we derive~\cite{Note0}
\begin{equation}\label{eq:erasure_fraction_bell}
r_\infty^\text{Bell} = \frac{\left\langle S_E(\ket{\phi_N})\right\rangle}{N\cdot 1\,\text{bit} } = 1 - N/L. 
\end{equation}
  This result is natural: The more states are excited, the more likely it is to occupy states of the same set of EEPS simultaneously, which then erase each others contribution to the EE. 

\begin{figure}
\centering
\includegraphics[width = 0.98\linewidth]{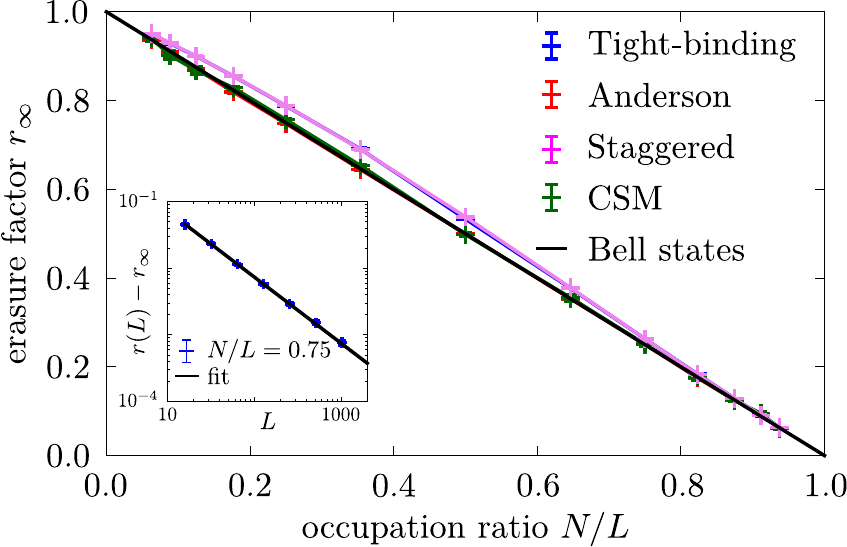}
\caption{Universal erasure factor $r_\infty$ at different occupation ratios $N/L$ for various single particle models. While the erasure of EE becomes negligible for low occupation ratios $N/L\ll 1$, entanglement erasing partner states are fully activated as $N/L\to 1$. All studied models follow the law $r_\infty \sim 1-N/L$, which is exact for a proposed model of Bell states (see text). The inset shows how we extrapolate our numerical data to the thermodynamic limit $L\to\infty$. Error bars are smaller than the dot sizes.}
\label{fig:erasure_factor}
\end{figure}

This thought can be directly applied to arbitrary models with very distinct structures of eigenstates. In \ref{fig:erasure_factor} we compare the erasure factors of different models and find a good agreement with our analytic result derived for the above constructed Bell states.  In our numerical tests, the expectation values of Eq.~\eqref{eq:erasurefactor} are taken over different combinations of jointly excited eigenstates and, if disorder is present, disorder ensembles. 
Beside the above discussed Anderson localized chain and the tight-binding model, we compare Eq.~\eqref{eq:erasure_fraction_bell} with randomly excited single particle states from the tight-binding Hamiltonian with staggered potentials, i.e. $h_i \to (-\mu)^i$, which creates a gap in the energy spectrum. Also, a central site model~\cite{Hetterich2017}, where an additional site $\ket{0}$ is coupled to each site of the above defined Anderson chain via the term $A/\sqrt{L}\sum_i ( c_i^\dagger c_0 + \text{h.c.})$, shows surprisingly good  agreement. This is remarkable because this central site model consists of a single particle mobility edge multifractal eigenstates~\cite{Hetterich2017}.

We hence find that all studied single particle models, despite their significant differences in the structure of their eigenstates, show a universal behavior of the joint~EE
\begin{align}
S_E(\ket{\phi_N}) &\approx \left(1-\frac{N}{L}\right) \sum_i^N S_E(\ket{e_i}).
\end{align}
Hence, the computation of the EE of a many-particle state $\ket{\phi_N}$, can be well-approximated by the filling ratio $N/L$ and the EE of single particle states, which is given in Eq.~\eqref{eq:singleparticleentropy}. This reduces the complexity of the computation from a diagonalization of matrices of the size of single particle Hilbert spaces to simple summations. 

\emph{Conclusion and Outlook ---} We have demonstrated that the excitation of additional states may reduce the total bipartite entanglement entropy of a system. This erasure of entanglement entropy may be used to completely uncorrelate different parts of a quantum system. Our analytic results for free models also show a  surprisingly well applicability for many-particle eigenstates of interacting models. 
The importance of controlling such quantum correlations, for instance in the initialization of states in quantum experiments, motivates to further examine our described mechanism of entanglement erasure in future works. Concretely we see the possibility of an explicit construction of entanglement erasing partner states for given models and a better understanding of entanglement erasure in interacting models. 

We thank  Felix Keidel, Frank Pollmann, Lorenzo Privitera, Benedikt Scharf, Adam Smith, Giuseppe De Tomasi and Bj\"orn Trauzettel for insightful discussions.

\bibliography{library}

\begin{thebibliography}{40}%
\makeatletter
\providecommand \@ifxundefined [1]{%
 \@ifx{#1\undefined}
}%
\providecommand \@ifnum [1]{%
 \ifnum #1\expandafter \@firstoftwo
 \else \expandafter \@secondoftwo
 \fi
}%
\providecommand \@ifx [1]{%
 \ifx #1\expandafter \@firstoftwo
 \else \expandafter \@secondoftwo
 \fi
}%
\providecommand \natexlab [1]{#1}%
\providecommand \enquote  [1]{``#1''}%
\providecommand \bibnamefont  [1]{#1}%
\providecommand \bibfnamefont [1]{#1}%
\providecommand \citenamefont [1]{#1}%
\providecommand \href@noop [0]{\@secondoftwo}%
\providecommand \href [0]{\begingroup \@sanitize@url \@href}%
\providecommand \@href[1]{\@@startlink{#1}\@@href}%
\providecommand \@@href[1]{\endgroup#1\@@endlink}%
\providecommand \@sanitize@url [0]{\catcode `\\12\catcode `\$12\catcode
  `\&12\catcode `\#12\catcode `\^12\catcode `\_12\catcode `\%12\relax}%
\providecommand \@@startlink[1]{}%
\providecommand \@@endlink[0]{}%
\providecommand \url  [0]{\begingroup\@sanitize@url \@url }%
\providecommand \@url [1]{\endgroup\@href {#1}{\urlprefix }}%
\providecommand \urlprefix  [0]{URL }%
\providecommand \Eprint [0]{\href }%
\providecommand \doibase [0]{http://dx.doi.org/}%
\providecommand \selectlanguage [0]{\@gobble}%
\providecommand \bibinfo  [0]{\@secondoftwo}%
\providecommand \bibfield  [0]{\@secondoftwo}%
\providecommand \translation [1]{[#1]}%
\providecommand \BibitemOpen [0]{}%
\providecommand \bibitemStop [0]{}%
\providecommand \bibitemNoStop [0]{.\EOS\space}%
\providecommand \EOS [0]{\spacefactor3000\relax}%
\providecommand \BibitemShut  [1]{\csname bibitem#1\endcsname}%
\let\auto@bib@innerbib\@empty
\bibitem [{\citenamefont {Einstein}\ \emph {et~al.}(1935)\citenamefont
  {Einstein}, \citenamefont {Podolsky},\ and\ \citenamefont
  {Rosen}}]{Einstein1935}%
  \BibitemOpen
  \bibfield  {author} {\bibinfo {author} {\bibfnamefont {A.}~\bibnamefont
  {Einstein}}, \bibinfo {author} {\bibfnamefont {B.}~\bibnamefont {Podolsky}},
  \ and\ \bibinfo {author} {\bibfnamefont {N.}~\bibnamefont {Rosen}},\
  }\href@noop {} {\bibfield  {journal} {\bibinfo  {journal} {Physical Review}\
  }\textbf {\bibinfo {volume} {47}},\ \bibinfo {pages} {777} (\bibinfo {year}
  {1935})}\BibitemShut {NoStop}%
\bibitem [{\citenamefont {Bell}(1964)}]{Bell1964}%
  \BibitemOpen
  \bibfield  {author} {\bibinfo {author} {\bibfnamefont {J.~S.}\ \bibnamefont
  {Bell}},\ }\href@noop {} {\bibfield  {journal} {\bibinfo  {journal}
  {Physics}\ }\textbf {\bibinfo {volume} {1}},\ \bibinfo {pages} {195}
  (\bibinfo {year} {1964})}\BibitemShut {NoStop}%
\bibitem [{\citenamefont {Bombelli}\ \emph {et~al.}(1986)\citenamefont
  {Bombelli}, \citenamefont {Koul}, \citenamefont {Lee},\ and\ \citenamefont
  {Sorkin}}]{Bombelli1986}%
  \BibitemOpen
  \bibfield  {author} {\bibinfo {author} {\bibfnamefont {L.}~\bibnamefont
  {Bombelli}}, \bibinfo {author} {\bibfnamefont {R.~K.}\ \bibnamefont {Koul}},
  \bibinfo {author} {\bibfnamefont {J.}~\bibnamefont {Lee}}, \ and\ \bibinfo
  {author} {\bibfnamefont {R.~D.}\ \bibnamefont {Sorkin}},\ }\href {\doibase
  10.1103/PhysRevD.34.373} {\bibfield  {journal} {\bibinfo  {journal} {Phys.
  Rev. D}\ }\textbf {\bibinfo {volume} {34}},\ \bibinfo {pages} {373} (\bibinfo
  {year} {1986})}\BibitemShut {NoStop}%
\bibitem [{\citenamefont {Srednicki}(1993)}]{Srednicki1993}%
  \BibitemOpen
  \bibfield  {author} {\bibinfo {author} {\bibfnamefont {M.}~\bibnamefont
  {Srednicki}},\ }\href {\doibase 10.1103/PhysRevLett.71.666} {\bibfield
  {journal} {\bibinfo  {journal} {Phys. Rev. Lett.}\ }\textbf {\bibinfo
  {volume} {71}},\ \bibinfo {pages} {666} (\bibinfo {year} {1993})}\BibitemShut
  {NoStop}%
\bibitem [{\citenamefont {Sarovar}\ \emph {et~al.}(2010)\citenamefont
  {Sarovar}, \citenamefont {Ishizaki}, \citenamefont {Fleming},\ and\
  \citenamefont {Whaley}}]{Sarovar2010}%
  \BibitemOpen
  \bibfield  {author} {\bibinfo {author} {\bibfnamefont {M.}~\bibnamefont
  {Sarovar}}, \bibinfo {author} {\bibfnamefont {A.}~\bibnamefont {Ishizaki}},
  \bibinfo {author} {\bibfnamefont {G.~R.}\ \bibnamefont {Fleming}}, \ and\
  \bibinfo {author} {\bibfnamefont {K.~B.}\ \bibnamefont {Whaley}},\ }\href
  {\doibase 10.1038/nphys1652} {\bibfield  {journal} {\bibinfo  {journal}
  {Nature Physics}\ }\textbf {\bibinfo {volume} {6}},\ \bibinfo {pages} {462}
  (\bibinfo {year} {2010})}\BibitemShut {NoStop}%
\bibitem [{\citenamefont {Nielsen}\ and\ \citenamefont
  {Chuang}(2011)}]{Nielsen2011}%
  \BibitemOpen
  \bibfield  {author} {\bibinfo {author} {\bibfnamefont {M.~A.}\ \bibnamefont
  {Nielsen}}\ and\ \bibinfo {author} {\bibfnamefont {I.~L.}\ \bibnamefont
  {Chuang}},\ }\href@noop {} {\emph {\bibinfo {title} {Quantum Computation and
  Quantum Information: 10th Anniversary Edition}}},\ \bibinfo {edition} {10th}\
  ed.\ (\bibinfo  {publisher} {Cambridge University Press},\ \bibinfo {address}
  {New York, NY, USA},\ \bibinfo {year} {2011})\BibitemShut {NoStop}%
\bibitem [{\citenamefont {Lieb}\ and\ \citenamefont
  {Robinson}(1972)}]{Lieb1972}%
  \BibitemOpen
  \bibfield  {author} {\bibinfo {author} {\bibfnamefont {E.~H.}\ \bibnamefont
  {Lieb}}\ and\ \bibinfo {author} {\bibfnamefont {D.~W.}\ \bibnamefont
  {Robinson}},\ }\href {\doibase 10.1007/BF01645779} {\bibfield  {journal}
  {\bibinfo  {journal} {Communications in Mathematical Physics}\ }\textbf
  {\bibinfo {volume} {28}},\ \bibinfo {pages} {251} (\bibinfo {year}
  {1972})}\BibitemShut {NoStop}%
\bibitem [{\citenamefont {Shi}(2003)}]{Shi2003}%
  \BibitemOpen
  \bibfield  {author} {\bibinfo {author} {\bibfnamefont {Y.}~\bibnamefont
  {Shi}},\ }\href {\doibase 10.1103/PhysRevA.67.024301} {\bibfield  {journal}
  {\bibinfo  {journal} {Physical Review A}\ }\textbf {\bibinfo {volume} {67}},\
  \bibinfo {pages} {024301} (\bibinfo {year} {2003})},\ \Eprint
  {http://arxiv.org/abs/0205069} {arXiv:0205069 [quant-ph]} \BibitemShut
  {NoStop}%
\bibitem [{\citenamefont {Amico}\ \emph {et~al.}(2008)\citenamefont {Amico},
  \citenamefont {Fazio}, \citenamefont {Osterloh},\ and\ \citenamefont
  {Vedral}}]{Amico2008}%
  \BibitemOpen
  \bibfield  {author} {\bibinfo {author} {\bibfnamefont {L.}~\bibnamefont
  {Amico}}, \bibinfo {author} {\bibfnamefont {R.}~\bibnamefont {Fazio}},
  \bibinfo {author} {\bibfnamefont {A.}~\bibnamefont {Osterloh}}, \ and\
  \bibinfo {author} {\bibfnamefont {V.}~\bibnamefont {Vedral}},\ }\href
  {\doibase 10.1103/RevModPhys.80.517} {\bibfield  {journal} {\bibinfo
  {journal} {Rev. Mod. Phys.}\ }\textbf {\bibinfo {volume} {80}},\ \bibinfo
  {pages} {517} (\bibinfo {year} {2008})}\BibitemShut {NoStop}%
\bibitem [{\citenamefont {Alba}\ \emph {et~al.}(2009)\citenamefont {Alba},
  \citenamefont {Fagotti},\ and\ \citenamefont {Calabrese}}]{Alba2009}%
  \BibitemOpen
  \bibfield  {author} {\bibinfo {author} {\bibfnamefont {V.}~\bibnamefont
  {Alba}}, \bibinfo {author} {\bibfnamefont {M.}~\bibnamefont {Fagotti}}, \
  and\ \bibinfo {author} {\bibfnamefont {P.}~\bibnamefont {Calabrese}},\ }\href
  {\doibase 10.1088/1742-5468/2009/10/P10020} {\bibfield  {journal} {\bibinfo
  {journal} {J. Stat. Mech. Theory Exp.}\ }\textbf {\bibinfo {volume} {2009}},\
  \bibinfo {pages} {P10020} (\bibinfo {year} {2009})}\BibitemShut {NoStop}%
\bibitem [{\citenamefont {Eisert}\ \emph {et~al.}(2010)\citenamefont {Eisert},
  \citenamefont {Cramer},\ and\ \citenamefont {Plenio}}]{Eisert2010}%
  \BibitemOpen
  \bibfield  {author} {\bibinfo {author} {\bibfnamefont {J.}~\bibnamefont
  {Eisert}}, \bibinfo {author} {\bibfnamefont {M.}~\bibnamefont {Cramer}}, \
  and\ \bibinfo {author} {\bibfnamefont {M.~B.}\ \bibnamefont {Plenio}},\
  }\href {\doibase 10.1103/RevModPhys.82.277} {\bibfield  {journal} {\bibinfo
  {journal} {Rev. Mod. Phys.}\ }\textbf {\bibinfo {volume} {82}},\ \bibinfo
  {pages} {277} (\bibinfo {year} {2010})}\BibitemShut {NoStop}%
\bibitem [{\citenamefont {Cheneau}\ \emph {et~al.}(2012)\citenamefont
  {Cheneau}, \citenamefont {Barmettler}, \citenamefont {Poletti}, \citenamefont
  {Endres}, \citenamefont {Schau{\ss}}, \citenamefont {Fukuhara}, \citenamefont
  {Gross}, \citenamefont {Bloch}, \citenamefont {Kollath},\ and\ \citenamefont
  {Kuhr}}]{Cheneau2012}%
  \BibitemOpen
  \bibfield  {author} {\bibinfo {author} {\bibfnamefont {M.}~\bibnamefont
  {Cheneau}}, \bibinfo {author} {\bibfnamefont {P.}~\bibnamefont {Barmettler}},
  \bibinfo {author} {\bibfnamefont {D.}~\bibnamefont {Poletti}}, \bibinfo
  {author} {\bibfnamefont {M.}~\bibnamefont {Endres}}, \bibinfo {author}
  {\bibfnamefont {P.}~\bibnamefont {Schau{\ss}}}, \bibinfo {author}
  {\bibfnamefont {T.}~\bibnamefont {Fukuhara}}, \bibinfo {author}
  {\bibfnamefont {C.}~\bibnamefont {Gross}}, \bibinfo {author} {\bibfnamefont
  {I.}~\bibnamefont {Bloch}}, \bibinfo {author} {\bibfnamefont
  {C.}~\bibnamefont {Kollath}}, \ and\ \bibinfo {author} {\bibfnamefont
  {S.}~\bibnamefont {Kuhr}},\ }\href {https://doi.org/10.1038/nature10748}
  {\bibfield  {journal} {\bibinfo  {journal} {Nature}\ }\textbf {\bibinfo
  {volume} {481}},\ \bibinfo {pages} {484 EP } (\bibinfo {year}
  {2012})}\BibitemShut {NoStop}%
\bibitem [{\citenamefont {Ganahl}\ \emph {et~al.}(2012)\citenamefont {Ganahl},
  \citenamefont {Rabel}, \citenamefont {Essler},\ and\ \citenamefont
  {Evertz}}]{Ganahl2012}%
  \BibitemOpen
  \bibfield  {author} {\bibinfo {author} {\bibfnamefont {M.}~\bibnamefont
  {Ganahl}}, \bibinfo {author} {\bibfnamefont {E.}~\bibnamefont {Rabel}},
  \bibinfo {author} {\bibfnamefont {F.~H.~L.}\ \bibnamefont {Essler}}, \ and\
  \bibinfo {author} {\bibfnamefont {H.~G.}\ \bibnamefont {Evertz}},\ }\href
  {\doibase 10.1103/PhysRevLett.108.077206} {\bibfield  {journal} {\bibinfo
  {journal} {Phys. Rev. Lett.}\ }\textbf {\bibinfo {volume} {108}},\ \bibinfo
  {pages} {077206} (\bibinfo {year} {2012})}\BibitemShut {NoStop}%
\bibitem [{\citenamefont {Igl\'oi}\ \emph {et~al.}(2012)\citenamefont
  {Igl\'oi}, \citenamefont {Szatm\'ari},\ and\ \citenamefont
  {Lin}}]{Ferenc2012}%
  \BibitemOpen
  \bibfield  {author} {\bibinfo {author} {\bibfnamefont {F.}~\bibnamefont
  {Igl\'oi}}, \bibinfo {author} {\bibfnamefont {Z.}~\bibnamefont {Szatm\'ari}},
  \ and\ \bibinfo {author} {\bibfnamefont {Y.-C.}\ \bibnamefont {Lin}},\ }\href
  {\doibase 10.1103/PhysRevB.85.094417} {\bibfield  {journal} {\bibinfo
  {journal} {Phys. Rev. B}\ }\textbf {\bibinfo {volume} {85}},\ \bibinfo
  {pages} {094417} (\bibinfo {year} {2012})}\BibitemShut {NoStop}%
\bibitem [{\citenamefont {Collura}\ and\ \citenamefont
  {Calabrese}(2013)}]{Collura2013}%
  \BibitemOpen
  \bibfield  {author} {\bibinfo {author} {\bibfnamefont {M.}~\bibnamefont
  {Collura}}\ and\ \bibinfo {author} {\bibfnamefont {P.}~\bibnamefont
  {Calabrese}},\ }\href {\doibase 10.1088/1751-8113/46/17/175001} {\bibfield
  {journal} {\bibinfo  {journal} {Journal of Physics A: Mathematical and
  Theoretical}\ }\textbf {\bibinfo {volume} {46}},\ \bibinfo {pages} {175001}
  (\bibinfo {year} {2013})}\BibitemShut {NoStop}%
\bibitem [{\citenamefont {de~Paula}\ \emph {et~al.}(2017)\citenamefont
  {de~Paula}, \citenamefont {Bragan\ifmmode~\mbox{\c{c}}\else \c{c}\fi{}a},
  \citenamefont {Pereira}, \citenamefont {Drumond},\ and\ \citenamefont
  {Aguiar}}]{Paula2017}%
  \BibitemOpen
  \bibfield  {author} {\bibinfo {author} {\bibfnamefont {A.~L.}\ \bibnamefont
  {de~Paula}}, \bibinfo {author} {\bibfnamefont {H.}~\bibnamefont
  {Bragan\ifmmode~\mbox{\c{c}}\else \c{c}\fi{}a}}, \bibinfo {author}
  {\bibfnamefont {R.~G.}\ \bibnamefont {Pereira}}, \bibinfo {author}
  {\bibfnamefont {R.~C.}\ \bibnamefont {Drumond}}, \ and\ \bibinfo {author}
  {\bibfnamefont {M.~C.~O.}\ \bibnamefont {Aguiar}},\ }\href {\doibase
  10.1103/PhysRevB.95.045125} {\bibfield  {journal} {\bibinfo  {journal} {Phys.
  Rev. B}\ }\textbf {\bibinfo {volume} {95}},\ \bibinfo {pages} {045125}
  (\bibinfo {year} {2017})}\BibitemShut {NoStop}%
\bibitem [{\citenamefont {\ifmmode \check{Z}\else
  \v{Z}\fi{}nidari\ifmmode~\check{c}\else \v{c}\fi{}}\ \emph
  {et~al.}(2008)\citenamefont {\ifmmode \check{Z}\else
  \v{Z}\fi{}nidari\ifmmode~\check{c}\else \v{c}\fi{}}, \citenamefont {Prosen},\
  and\ \citenamefont {Prelov\ifmmode~\check{s}\else
  \v{s}\fi{}ek}}]{Znidaric2008}%
  \BibitemOpen
  \bibfield  {author} {\bibinfo {author} {\bibfnamefont {M.}~\bibnamefont
  {\ifmmode \check{Z}\else \v{Z}\fi{}nidari\ifmmode~\check{c}\else
  \v{c}\fi{}}}, \bibinfo {author} {\bibfnamefont {T.~c.~v.}\ \bibnamefont
  {Prosen}}, \ and\ \bibinfo {author} {\bibfnamefont {P.}~\bibnamefont
  {Prelov\ifmmode~\check{s}\else \v{s}\fi{}ek}},\ }\href {\doibase
  10.1103/PhysRevB.77.064426} {\bibfield  {journal} {\bibinfo  {journal} {Phys.
  Rev. B}\ }\textbf {\bibinfo {volume} {77}},\ \bibinfo {pages} {064426}
  (\bibinfo {year} {2008})}\BibitemShut {NoStop}%
\bibitem [{\citenamefont {Bardarson}\ \emph {et~al.}(2012)\citenamefont
  {Bardarson}, \citenamefont {Pollmann},\ and\ \citenamefont
  {Moore}}]{Bardarson2012}%
  \BibitemOpen
  \bibfield  {author} {\bibinfo {author} {\bibfnamefont {J.~H.}\ \bibnamefont
  {Bardarson}}, \bibinfo {author} {\bibfnamefont {F.}~\bibnamefont {Pollmann}},
  \ and\ \bibinfo {author} {\bibfnamefont {J.~E.}\ \bibnamefont {Moore}},\
  }\href {\doibase 10.1103/PhysRevLett.109.017202} {\bibfield  {journal}
  {\bibinfo  {journal} {Phys. Rev. Lett.}\ }\textbf {\bibinfo {volume} {109}},\
  \bibinfo {pages} {017202} (\bibinfo {year} {2012})}\BibitemShut {NoStop}%
\bibitem [{\citenamefont {Serbyn}\ \emph {et~al.}(2014)\citenamefont {Serbyn},
  \citenamefont {Papi\ifmmode~\acute{c}\else \'{c}\fi{}},\ and\ \citenamefont
  {Abanin}}]{Serbyn2014}%
  \BibitemOpen
  \bibfield  {author} {\bibinfo {author} {\bibfnamefont {M.}~\bibnamefont
  {Serbyn}}, \bibinfo {author} {\bibfnamefont {Z.}~\bibnamefont
  {Papi\ifmmode~\acute{c}\else \'{c}\fi{}}}, \ and\ \bibinfo {author}
  {\bibfnamefont {D.~A.}\ \bibnamefont {Abanin}},\ }\href {\doibase
  10.1103/PhysRevB.90.174302} {\bibfield  {journal} {\bibinfo  {journal} {Phys.
  Rev. B}\ }\textbf {\bibinfo {volume} {90}},\ \bibinfo {pages} {174302}
  (\bibinfo {year} {2014})}\BibitemShut {NoStop}%
\bibitem [{\citenamefont {Hetterich}\ \emph {et~al.}(2017)\citenamefont
  {Hetterich}, \citenamefont {Serbyn}, \citenamefont {Dom{\'{i}}nguez},
  \citenamefont {Pollmann},\ and\ \citenamefont {Trauzettel}}]{Hetterich2017}%
  \BibitemOpen
  \bibfield  {author} {\bibinfo {author} {\bibfnamefont {D.}~\bibnamefont
  {Hetterich}}, \bibinfo {author} {\bibfnamefont {M.}~\bibnamefont {Serbyn}},
  \bibinfo {author} {\bibfnamefont {F.}~\bibnamefont {Dom{\'{i}}nguez}},
  \bibinfo {author} {\bibfnamefont {F.}~\bibnamefont {Pollmann}}, \ and\
  \bibinfo {author} {\bibfnamefont {B.}~\bibnamefont {Trauzettel}},\ }\href
  {\doibase 10.1103/PhysRevB.96.104203} {\bibfield  {journal} {\bibinfo
  {journal} {Phys. Rev. B}\ }\textbf {\bibinfo {volume} {96}},\ \bibinfo
  {pages} {104203} (\bibinfo {year} {2017})}\BibitemShut {NoStop}%
\bibitem [{\citenamefont {De~Tomasi}\ \emph {et~al.}(2017)\citenamefont
  {De~Tomasi}, \citenamefont {Bera}, \citenamefont {Bardarson},\ and\
  \citenamefont {Pollmann}}]{DeTomasi2017}%
  \BibitemOpen
  \bibfield  {author} {\bibinfo {author} {\bibfnamefont {G.}~\bibnamefont
  {De~Tomasi}}, \bibinfo {author} {\bibfnamefont {S.}~\bibnamefont {Bera}},
  \bibinfo {author} {\bibfnamefont {J.~H.}\ \bibnamefont {Bardarson}}, \ and\
  \bibinfo {author} {\bibfnamefont {F.}~\bibnamefont {Pollmann}},\ }\href
  {\doibase 10.1103/PhysRevLett.118.016804} {\bibfield  {journal} {\bibinfo
  {journal} {Phys. Rev. Lett.}\ }\textbf {\bibinfo {volume} {118}},\ \bibinfo
  {pages} {016804} (\bibinfo {year} {2017})}\BibitemShut {NoStop}%
\bibitem [{\citenamefont {Smith}\ \emph {et~al.}(2017)\citenamefont {Smith},
  \citenamefont {Knolle}, \citenamefont {Kovrizhin},\ and\ \citenamefont
  {Moessner}}]{Smith2017}%
  \BibitemOpen
  \bibfield  {author} {\bibinfo {author} {\bibfnamefont {A.}~\bibnamefont
  {Smith}}, \bibinfo {author} {\bibfnamefont {J.}~\bibnamefont {Knolle}},
  \bibinfo {author} {\bibfnamefont {D.~L.}\ \bibnamefont {Kovrizhin}}, \ and\
  \bibinfo {author} {\bibfnamefont {R.}~\bibnamefont {Moessner}},\ }\href
  {\doibase 10.1103/PhysRevLett.118.266601} {\bibfield  {journal} {\bibinfo
  {journal} {Phys. Rev. Lett.}\ }\textbf {\bibinfo {volume} {118}},\ \bibinfo
  {pages} {266601} (\bibinfo {year} {2017})}\BibitemShut {NoStop}%
\bibitem [{\citenamefont {Berkovits}(2013)}]{Berkovits2013}%
  \BibitemOpen
  \bibfield  {author} {\bibinfo {author} {\bibfnamefont {R.}~\bibnamefont
  {Berkovits}},\ }\href {\doibase 10.1103/PhysRevB.87.075141} {\bibfield
  {journal} {\bibinfo  {journal} {Phys. Rev. B}\ }\textbf {\bibinfo {volume}
  {87}},\ \bibinfo {pages} {075141} (\bibinfo {year} {2013})}\BibitemShut
  {NoStop}%
\bibitem [{\citenamefont {M{\"{o}}lter}\ \emph {et~al.}(2014)\citenamefont
  {M{\"{o}}lter}, \citenamefont {Barthel}, \citenamefont {Schollw{\"{o}}ck},\
  and\ \citenamefont {Alba}}]{Molter2014}%
  \BibitemOpen
  \bibfield  {author} {\bibinfo {author} {\bibfnamefont {J.}~\bibnamefont
  {M{\"{o}}lter}}, \bibinfo {author} {\bibfnamefont {T.}~\bibnamefont
  {Barthel}}, \bibinfo {author} {\bibfnamefont {U.}~\bibnamefont
  {Schollw{\"{o}}ck}}, \ and\ \bibinfo {author} {\bibfnamefont
  {V.}~\bibnamefont {Alba}},\ }\href {\doibase
  10.1088/1742-5468/2014/10/P10029} {\bibfield  {journal} {\bibinfo  {journal}
  {J. Stat. Mech. Theory Exp.}\ }\textbf {\bibinfo {volume} {2014}},\ \bibinfo
  {pages} {P10029} (\bibinfo {year} {2014})}\BibitemShut {NoStop}%
\bibitem [{\citenamefont {Storms}\ and\ \citenamefont
  {Singh}(2014)}]{Storms2014}%
  \BibitemOpen
  \bibfield  {author} {\bibinfo {author} {\bibfnamefont {M.}~\bibnamefont
  {Storms}}\ and\ \bibinfo {author} {\bibfnamefont {R.~R.~P.}\ \bibnamefont
  {Singh}},\ }\href {\doibase 10.1103/PhysRevE.89.012125} {\bibfield  {journal}
  {\bibinfo  {journal} {Phys. Rev. E}\ }\textbf {\bibinfo {volume} {89}},\
  \bibinfo {pages} {012125} (\bibinfo {year} {2014})}\BibitemShut {NoStop}%
\bibitem [{\citenamefont {Ramírez}\ \emph {et~al.}(2014)\citenamefont
  {Ramírez}, \citenamefont {Rodríguez-Laguna},\ and\ \citenamefont
  {Sierra}}]{Ramarez2014}%
  \BibitemOpen
  \bibfield  {author} {\bibinfo {author} {\bibfnamefont {G.}~\bibnamefont
  {Ramírez}}, \bibinfo {author} {\bibfnamefont {J.}~\bibnamefont
  {Rodríguez-Laguna}}, \ and\ \bibinfo {author} {\bibfnamefont
  {G.}~\bibnamefont {Sierra}},\ }\href
  {http://stacks.iop.org/1742-5468/2014/i=7/a=P07003} {\bibfield  {journal}
  {\bibinfo  {journal} {Journal of Statistical Mechanics: Theory and
  Experiment}\ }\textbf {\bibinfo {volume} {2014}},\ \bibinfo {pages} {P07003}
  (\bibinfo {year} {2014})}\BibitemShut {NoStop}%
\bibitem [{\citenamefont {Pizorn}(2012)}]{Pizorn2012}%
  \BibitemOpen
  \bibfield  {author} {\bibinfo {author} {\bibfnamefont {I.}~\bibnamefont
  {Pizorn}},\ }\href {https://arxiv.org/pdf/1202.3336.pdf} {\  (\bibinfo {year}
  {2012})},\ \Eprint {http://arxiv.org/abs/1202.3336} {arXiv:1202.3336}
  \BibitemShut {NoStop}%
\bibitem [{\citenamefont {Castro-Alvaredo}\ \emph {et~al.}(2018)\citenamefont
  {Castro-Alvaredo}, \citenamefont {De~Fazio}, \citenamefont {Doyon},\ and\
  \citenamefont {Sz\'ecs\'enyi}}]{CastroAlvaredo2018}%
  \BibitemOpen
  \bibfield  {author} {\bibinfo {author} {\bibfnamefont {O.~A.}\ \bibnamefont
  {Castro-Alvaredo}}, \bibinfo {author} {\bibfnamefont {C.}~\bibnamefont
  {De~Fazio}}, \bibinfo {author} {\bibfnamefont {B.}~\bibnamefont {Doyon}}, \
  and\ \bibinfo {author} {\bibfnamefont {I.~M.}\ \bibnamefont
  {Sz\'ecs\'enyi}},\ }\href {\doibase 10.1103/PhysRevLett.121.170602}
  {\bibfield  {journal} {\bibinfo  {journal} {Phys. Rev. Lett.}\ }\textbf
  {\bibinfo {volume} {121}},\ \bibinfo {pages} {170602} (\bibinfo {year}
  {2018})}\BibitemShut {NoStop}%
\bibitem [{\citenamefont {Araki}\ and\ \citenamefont {Lieb}(1970)}]{araki1970}%
  \BibitemOpen
  \bibfield  {author} {\bibinfo {author} {\bibfnamefont {H.}~\bibnamefont
  {Araki}}\ and\ \bibinfo {author} {\bibfnamefont {E.~H.}\ \bibnamefont
  {Lieb}},\ }\href {https://projecteuclid.org:443/euclid.cmp/1103842506}
  {\bibfield  {journal} {\bibinfo  {journal} {Comm. Math. Phys.}\ }\textbf
  {\bibinfo {volume} {18}},\ \bibinfo {pages} {160} (\bibinfo {year}
  {1970})}\BibitemShut {NoStop}%
\bibitem [{\citenamefont {Peschel}\ and\ \citenamefont
  {Eisler}(2009)}]{Peschel2009}%
  \BibitemOpen
  \bibfield  {author} {\bibinfo {author} {\bibfnamefont {I.}~\bibnamefont
  {Peschel}}\ and\ \bibinfo {author} {\bibfnamefont {V.}~\bibnamefont
  {Eisler}},\ }\href {http://stacks.iop.org/1751-8121/42/i=50/a=504003}
  {\bibfield  {journal} {\bibinfo  {journal} {Journal of Physics A:
  Mathematical and Theoretical}\ }\textbf {\bibinfo {volume} {42}},\ \bibinfo
  {pages} {504003} (\bibinfo {year} {2009})}\BibitemShut {NoStop}%
\bibitem [{Note0()}]{Note0}%
  \BibitemOpen
  \bibinfo {note} {See supplementary material for the derivation of the proof,
  the derivation of the erasure, and details on the numerical
  procedures.}\BibitemShut {Stop}%
\bibitem [{\citenamefont {Linden}\ \emph {et~al.}(2006)\citenamefont {Linden},
  \citenamefont {Popescu},\ and\ \citenamefont {Smolin}}]{Linden2006}%
  \BibitemOpen
  \bibfield  {author} {\bibinfo {author} {\bibfnamefont {N.}~\bibnamefont
  {Linden}}, \bibinfo {author} {\bibfnamefont {S.}~\bibnamefont {Popescu}}, \
  and\ \bibinfo {author} {\bibfnamefont {J.~A.}\ \bibnamefont {Smolin}},\
  }\href {\doibase 10.1103/PhysRevLett.97.100502} {\bibfield  {journal}
  {\bibinfo  {journal} {Physical Review Letters}\ }\textbf {\bibinfo {volume}
  {97}},\ \bibinfo {pages} {100502} (\bibinfo {year} {2006})},\ \Eprint
  {http://arxiv.org/abs/0507049v2} {arXiv:0507049v2 [arXiv:quant-ph]}
  \BibitemShut {NoStop}%
\bibitem [{\citenamefont {Anderson}(1958)}]{Anderson1958}%
  \BibitemOpen
  \bibfield  {author} {\bibinfo {author} {\bibfnamefont {P.~W.}\ \bibnamefont
  {Anderson}},\ }\href {\doibase 10.1103/PhysRev.109.1492} {\bibfield
  {journal} {\bibinfo  {journal} {Phys. Rev.}\ }\textbf {\bibinfo {volume}
  {109}},\ \bibinfo {pages} {1492} (\bibinfo {year} {1958})},\ \Eprint
  {http://arxiv.org/abs/0807.2531} {arXiv:0807.2531} \BibitemShut {NoStop}%
\bibitem [{\citenamefont {Abrahams}\ \emph {et~al.}(1979)\citenamefont
  {Abrahams}, \citenamefont {Anderson}, \citenamefont {Licciardello},\ and\
  \citenamefont {Ramakrishnan}}]{Anderson1979}%
  \BibitemOpen
  \bibfield  {author} {\bibinfo {author} {\bibfnamefont {E.}~\bibnamefont
  {Abrahams}}, \bibinfo {author} {\bibfnamefont {P.~W.}\ \bibnamefont
  {Anderson}}, \bibinfo {author} {\bibfnamefont {D.~C.}\ \bibnamefont
  {Licciardello}}, \ and\ \bibinfo {author} {\bibfnamefont {T.~V.}\
  \bibnamefont {Ramakrishnan}},\ }\href {\doibase 10.1103/PhysRevLett.42.673}
  {\bibfield  {journal} {\bibinfo  {journal} {Phys. Rev. Lett.}\ }\textbf
  {\bibinfo {volume} {42}},\ \bibinfo {pages} {673} (\bibinfo {year}
  {1979})}\BibitemShut {NoStop}%
\bibitem [{\citenamefont {Mott}\ and\ \citenamefont {Twose}(1961)}]{Mott1961}%
  \BibitemOpen
  \bibfield  {author} {\bibinfo {author} {\bibfnamefont {N.}~\bibnamefont
  {Mott}}\ and\ \bibinfo {author} {\bibfnamefont {W.}~\bibnamefont {Twose}},\
  }\href {\doibase 10.1080/00018736100101271} {\bibfield  {journal} {\bibinfo
  {journal} {Advances in Physics}\ }\textbf {\bibinfo {volume} {10}},\ \bibinfo
  {pages} {107} (\bibinfo {year} {1961})},\ \Eprint
  {http://arxiv.org/abs/https://doi.org/10.1080/00018736100101271}
  {https://doi.org/10.1080/00018736100101271} \BibitemShut {NoStop}%
\bibitem [{\citenamefont {Basko}\ \emph {et~al.}(2006)\citenamefont {Basko},
  \citenamefont {Aleiner},\ and\ \citenamefont {Altshuler}}]{Basko2006a}%
  \BibitemOpen
  \bibfield  {author} {\bibinfo {author} {\bibfnamefont {D.}~\bibnamefont
  {Basko}}, \bibinfo {author} {\bibfnamefont {I.}~\bibnamefont {Aleiner}}, \
  and\ \bibinfo {author} {\bibfnamefont {B.}~\bibnamefont {Altshuler}},\ }\href
  {\doibase 10.1016/j.aop.2005.11.014} {\bibfield  {journal} {\bibinfo
  {journal} {Ann. Phys.}\ }\textbf {\bibinfo {volume} {321}},\ \bibinfo {pages}
  {1126} (\bibinfo {year} {2006})},\ \Eprint {http://arxiv.org/abs/0506617}
  {arXiv:0506617 [cond-mat]} \BibitemShut {NoStop}%
\bibitem [{\citenamefont {Oganesyan}\ and\ \citenamefont
  {Huse}(2007)}]{Oganesyan2007}%
  \BibitemOpen
  \bibfield  {author} {\bibinfo {author} {\bibfnamefont {V.}~\bibnamefont
  {Oganesyan}}\ and\ \bibinfo {author} {\bibfnamefont {D.~A.}\ \bibnamefont
  {Huse}},\ }\href {\doibase 10.1103/PhysRevB.75.155111} {\bibfield  {journal}
  {\bibinfo  {journal} {Phys. Rev. B}\ }\textbf {\bibinfo {volume} {75}},\
  \bibinfo {pages} {155111} (\bibinfo {year} {2007})},\ \Eprint
  {http://arxiv.org/abs/0610854} {arXiv:0610854 [cond-mat]} \BibitemShut
  {NoStop}%
\bibitem [{\citenamefont {Pal}\ and\ \citenamefont {Huse}(2010)}]{Pal2010}%
  \BibitemOpen
  \bibfield  {author} {\bibinfo {author} {\bibfnamefont {A.}~\bibnamefont
  {Pal}}\ and\ \bibinfo {author} {\bibfnamefont {D.~A.}\ \bibnamefont {Huse}},\
  }\href {\doibase 10.1103/PhysRevB.82.174411} {\bibfield  {journal} {\bibinfo
  {journal} {Phys. Rev. B}\ }\textbf {\bibinfo {volume} {82}},\ \bibinfo
  {pages} {174411} (\bibinfo {year} {2010})},\ \Eprint
  {http://arxiv.org/abs/1010.1992} {arXiv:1010.1992} \BibitemShut {NoStop}%
\bibitem [{\citenamefont {Luitz}\ \emph {et~al.}(2015)\citenamefont {Luitz},
  \citenamefont {Laflorencie},\ and\ \citenamefont {Alet}}]{Luitz2015}%
  \BibitemOpen
  \bibfield  {author} {\bibinfo {author} {\bibfnamefont {D.~J.}\ \bibnamefont
  {Luitz}}, \bibinfo {author} {\bibfnamefont {N.}~\bibnamefont {Laflorencie}},
  \ and\ \bibinfo {author} {\bibfnamefont {F.}~\bibnamefont {Alet}},\ }\href
  {\doibase 10.1103/PhysRevB.91.081103} {\bibfield  {journal} {\bibinfo
  {journal} {Phys. Rev. B}\ }\textbf {\bibinfo {volume} {91}},\ \bibinfo
  {pages} {081103(R)} (\bibinfo {year} {2015})},\ \Eprint
  {http://arxiv.org/abs/1411.0660} {arXiv:1411.0660} \BibitemShut {NoStop}%
\bibitem [{\citenamefont {Gogolin}\ \emph {et~al.}(2004)\citenamefont
  {Gogolin}, \citenamefont {Nersesian},\ and\ \citenamefont
  {Tsvelik}}]{Gogolin}%
  \BibitemOpen
  \bibfield  {author} {\bibinfo {author} {\bibfnamefont {A.~O.}\ \bibnamefont
  {Gogolin}}, \bibinfo {author} {\bibfnamefont {A.~A.}\ \bibnamefont
  {Nersesian}}, \ and\ \bibinfo {author} {\bibfnamefont {A.~M.}\ \bibnamefont
  {Tsvelik}},\ }\href@noop {} {\emph {\bibinfo {title} {{Bosonization and
  strongly correlated systems}}}}\ (\bibinfo {year} {2004})\BibitemShut
  {NoStop}%
\end{thebibliography}%

\appendix

\section{Proof of the subadditivity}
Here we prove the subadditivity of the entanglement entropy for two-particle excitations. In particular, given two orthonormal states $e_1^\dagger\ket{}, e_2^\dagger\ket{}$, where $e_i^\dagger$ excites a fermion in the state $i$ and $\ket{}$ is the vacuum state, the von Neumann entanglement entropy with respect to a bipartition $\mathcal{H}=\mathcal{H}_A\otimes \mathcal{H}_B$  satisfies the inequality:
\begin{equation}
 S_E(e_2^\dagger e_1^\dagger\ket{}) \leq  S_E(e_1^\dagger\ket{}) + S_E(e_2^\dagger\ket{}) 
\end{equation}

\emph{Proof ---} The von Neumann entanglement entropy of a state $\ket{\psi}$ is defined by 
\begin{equation}
S_E( \ket{\psi}) = - \tr{ \rho_A \ln \rho_A},
\end{equation}
where $\rho_A = \tr[B]{ \ket{\psi}\bra{\psi}}$ and $\tr[B]{\,}$ is the partial trace over $\mathcal{H}_B$. For Slater determinants $\ket{\psi} = e_{i_1}^\dagger \ldots e_{i_N}^\dagger\ket{\,}$ of $N$ single particle states  $S_E(\ket{\psi})$ can be evaluated by~\cite{Peschel2009}
\begin{equation}
S_E(\ket{\psi}) = - \tr{ C_A \ln C_A + (\tone-C_A) \ln (\tone-C_A)},
\end{equation}
where 
\begin{equation}
\label{corrM}
(C_A)_{ij} = \bra{\psi} {c_j^A}^\dagger c_i^A \ket{\psi}
\end{equation}
is a $(\dim \mathcal{H}_A) \times (\dim \mathcal{H}_A)$ correlation matrix with respect to a single particle basis $\{ {c_i^A}^\dagger \ket{\,}\}$ that spans $\mathcal{H}_A$. 
In order to conduct the proof we first derive the following two lemmas.
\\

\emph{\textbf{Lemma 1:}} The correlation matrices are additive with respect to a joint excitation, i.e. 
\begin{equation}
C^A(\ket{e_1e_2}) = C^A(\ket{e_1}) + C^A(\ket{e_2}).
\end{equation}

\emph{\textbf{Lemma 2:}} The correlation matrix of a single excitation $\ket{e_n}$ has at most one nonzero eigenvalue and can be expressed by
\begin{equation}\label{eq_lemma2}
C^A(\ket{e_n}) = \lambda_n \vec{\phi}_n (\vec{\phi}_n)^\dagger,
\end{equation}
where $\lambda_n$ is the probability of $\ket{e_n} = \sum\limits_i U_{ni} \ket{c_i}$  to be in $\mathcal{H}_A$, and the components of $\vec{\phi}_n$ are given by
\begin{equation}
\label{phi}
\left(\vec{\phi}_n\right)_i = \frac{1}{\sqrt{\lambda_n}}  U_{ni}.
\end{equation}

\emph{Proof of Lemma 1:} We employ a single particle basis $\{c_k^\dagger\ket{\,}\}$ of $\mathcal{H}$ that can be divided into two sets $\{{c_k^A}^\dagger\ket{\,}\}$ and $\{{c_k^B}^\dagger\ket{\,}\}$ that span $\mathcal{H}_A$ and $\mathcal{H}_B$, respectively. Because of the orthogonality of the two states $\ket{e_1}$ and $\ket{e_2}$, there exists a unitary matrix $U_{ik}$ such that
\begin{equation}
\ket{e_i} = \sum_k U_{ik} \ket{c_k}.
\end{equation}
The inverse of $U$ is given by $U^{-1}=U^\dagger$, therefore follows
   $c_k^\dagger = \sum\limits_l (U^\dagger)_{kl} e_l^\dagger$ and $c_k = \sum\limits_l (U^\dagger)^*_{kl} e_l= \sum\limits_l U_{lk} e_l$. Thus,
\begin{align}
C_{ij}(\ket{e_1e_2}) &= \bra{e_1e_2} c_j^\dagger c_i \ket{e_1e_2} \nonumber\\
&= \sum_{kl} (U^\dagger)_{jk} \bra{e_1e_2} e_k^\dagger e_l \ket{e_1e_2} U_{li} \nonumber\\
&= \sum_k (U^\dagger)_{jk} \bra{e_1e_2} e_k^\dagger e_k \ket{e_1e_2} U_{ki},\nonumber \\
\text{i.e. }C(\ket{e_1e_2}) &= U^\dagger D U,
\end{align}
where $D$ is diagonal, $D=\text{diag}(1,1,0,0,0,\ldots)$. The correlation matrix of the individual single particle excitations is analogously given by 
\begin{align}
C(\ket{e_1}) &= U^\dagger D_1 U \quad\text{with\;} D_1=\text{diag}(1,0,0,0,\ldots)  \nonumber\\
C(\ket{e_2}) &= U^\dagger D_2 U \quad\text{with\;} D_2=\text{diag}(0,1,0,0,\ldots) 
\end{align}  
with the same unitary matrix $U$ and, thus,  $D_1+D_2=D$. Hence,
\begin{align}
C(\ket{e_1e_2}) &= U^\dagger(D_1 + D_2) U  \nonumber\\
&= U^\dagger D_1 U + U^\dagger D_2 U \nonumber\\
&= C(\ket{e_1}) + C(\ket{e_2}),
\end{align}
and thus $C_{ij}(\ket{e_1e_2}) = C_{ij}(\ket{e_1}) + C_{ij}(\ket{e_2})$. As this holds for all indices $i,j$, this equation also holds for those values of $i,j$ that define the subspace $\mathcal{H}_A$, i.e. \begin{equation}
C^A(\ket{e_1e_2}) = C^A(\ket{e_1}) + C^A(\ket{e_2}).
\end{equation}

\emph{Proof of Lemma 2:} Using the definition of the correlation matrix and $\ket{e_n} = \sum\limits_i U_{ni} \ket{c_i}$ , it is simple to show that
\begin{align}
C^A_{ij}(\ket{e_n}) &=  \bra{e_n} {c_j^A}^\dagger c_i^A \ket{e_n}\\
&= \sum_{k,l} U_{nk} U_{nl}^* \bra{c_l}{c_j^A}^\dagger c_i^A \ket{e_n}\\
&= \left. U_{nj}^* U_{ni} \right\vert_{i,j\in A} \label{eq:partprovea}
\end{align}
Here, $U_{n,i}$ with fixed index $n$ represents a vector 
\begin{equation}
U_{n,i} = \left(\vec{\varphi}_n\right)_i
\end{equation}
that is \emph{not} normalized because $i$ is restricted to $i\in A$, which defines the subspace $\mathcal{H}_A$. We can thus define the normalized vector 
\begin{equation}
\left(\vec{\phi}_n\right)_i=  \frac{U_{ni}}{\sqrt{\lambda_n}} =  \frac{U_{ni}}{ \sqrt{\sum\limits_{j\in A} \abs{U_{nj}}^2}},
\end{equation} 
where $\lambda_n$ gives the squared overlap of the state $\ket{e_n}$ with subspace $\mathcal{H}_A$. Inserting into Eq.~\eqref{eq:partprovea} proves Lemma 2,
\begin{equation}
C^A(\ket{e_n}) = \lambda_n \vec{\phi}_n (\vec{\phi}_n)^\dagger,
\end{equation}
\\[1em]

\begin{figure}
\centering
\includegraphics[width=0.99\linewidth]{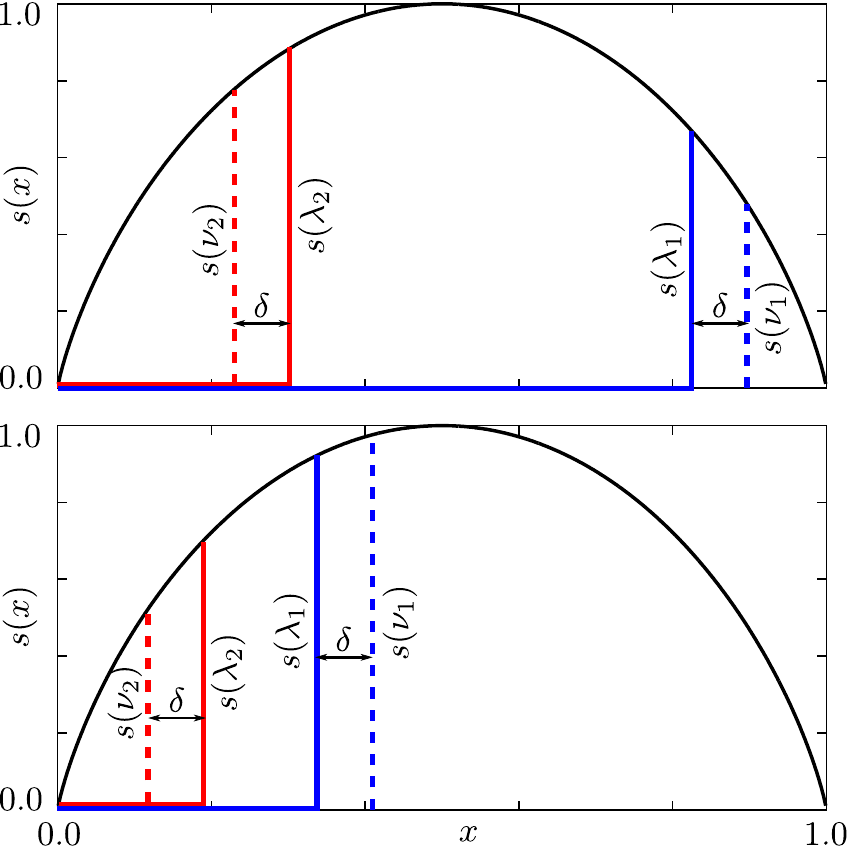}
\caption{Modification of the entanglement entropy at finite $\delta$. Top: For $\lambda_1$ and $\lambda_2$ on different sides of $x=0.5$ the reduction of the entanglement entropy is trivial. Bottom: For $\lambda_1, \lambda_2$ on the same side, a shift $\nu_2=\lambda_2 -\delta$, $\nu_1=\lambda_1+\delta$ always yields $s(\nu_2)+s(\nu_1) < s(\lambda_2) + s(\lambda_1)$ due to the concavity of $s(x)$. }
\label{fig:sofx}
\end{figure}

\emph{\textbf{Completion of the proof:}} Using Lemma 1 and Lemma 2, the reduced correlation matrix of the joint state $\ket{e_1e_2}$ is simply given by a superposition of  two projectors $\vec{\phi}_n (\vec{\phi}_n)^\dagger$, 
\begin{equation}
C^A(\ket{e_1e_2}) = \lambda_1 \vec{\phi}_1 (\vec{\phi}_1)^\dagger + \lambda_2 \vec{\phi}_2 (\vec{\phi}_2)^\dagger.
\end{equation}
Note that while the two rows $U_{1,i}, U_{2,i}$ of the unitary matrix $U$ are orthogonal to each other, this orthogonality disappears once one restricts $i$ to values $i\in A$. Hence, the vectors $\vec{\phi}_n$ are in general \emph{not} orthogonal to each other. Instead, the skalar product $\abs{\vec{\phi}_1 \vec{\phi}_2}^2$ quantifies the overlap of the two states $\ket{e_1},\ket{e_2}$ \emph{within} the subspace $\mathcal{H}_A$. 
Without loss of generality $\lambda_1>\lambda_2$. Then, the eigenvalues of $C^A$  yield 
\begin{align}
\nu_1 &= \lambda_1 + \delta/2, \label{eq_nu1}\\
\nu_2 &= \lambda_2 - \delta/2 \quad \text{with}\label{eq_nu2}\\
\delta &=  \sqrt{(\lambda_1-\lambda_2)^2 + 4\lambda_1\lambda_2 \abs{ \vec{\phi}_1 \vec{\phi}_2}^2} - (\lambda_1-\lambda_2) \label{eq_delta},
\end{align}
where $\delta=0$ if the scalar product   $\vec{\phi}_1\vec{\phi}_2$ vanishes  and $\delta>0$ elsewhere. 
As the function 
\begin{equation}
s(x) = - x \log_2 x - (1-x)\log_2(1-x)
\end{equation}
is concave, $\delta>0$ yields (see Fig.~\ref{fig:sofx} for an illustration)
\begin{align}
S_E(\ket{e_1 e_2}) &= s(\nu_1) + s(\nu_2) \\
&< s(\lambda_1) + s(\lambda_2) \\
&= S_E(\ket{e_1}) + S_E(\ket{E_2}).
\end{align}
The entanglement entropy is thus additive if and only if $\vec{\phi}_1\vec{\phi}_2=0$, taking aside trivial cases in which one $\lambda_i$ is zero or unity. Herewith the theorem is proven. 

\section{Entanglement entropy in tight-binding model}
Here we derive the EE for the two-particle eigenstate in the tight-binding model, given by the Eqs. (10) and (11) of the main text.  
The full correlation matrix of the two-particle eigenstate 
\begin{equation}
\label{exc_state}
|\Psi \rangle =|k_1k_2 \rangle= |k_1\rangle \otimes |k_2 \rangle,
\end{equation}
where $\ket{k} = \frac{1}{\sqrt{L}} \sum_{j=1}^L e^{i k j} c_j^\dagger \ket{}$
is then, according to Eq.~\eqref{corrM}, given by
\begin{align}
\label{corr_matrix_tb}
C_{ij}(|k_1 k_2\rangle) &=\frac{1}{L} \left[e^{ \mathrm{i} k_1(j-i)}+ e^{ \mathrm{i} k_2(j-i)} \right]\\
&=C_{ij}(|k_1 \rangle)+C_{ij}(|k_2 \rangle).
\end{align}

Consider now a contiguous bipartition $A$ and $B$ of the chain and study part $A$ of the size $L_A$. The reduced correlation function for this state is given by the corresponding block of the matrix $C_{ij}(|k_1 k_2\rangle)$, i.e. 
\begin{align}
\label{corr_submatrix_tb}
C^A_{ij}(|k_1k_2 \rangle)=C_{ij}(|k_1 k_2\rangle), \quad \text{with}\;i,j \in A
\end{align}
The normalized eigenvectors, defined in Eq.~\eqref{phi}, and eigenvalues of the reduced single particle correlation matrices $C_{ij}^A(|k_1 \rangle)$ and  $C_{ij}^A(|k_2 \rangle)$ yield 
\begin{align}
\label{eigensystem_A}
\lambda_1=\lambda_2&=\frac{L_A}{L},  \\
\left(\vec{\phi}_1\right)_j&=\frac{1}{\sqrt{L_A}} e^{-\mathrm{i} k_1j}, \\
\left(\vec{\phi}_2\right)_j&=\frac{1}{\sqrt{L_A}} e^{-\mathrm{i} k_2j} \;\text{with}\;  j \in A.
 \end{align}
Now we study subadditivity of the entanglement entropy. The eigenvalues of the two particle reduced correlation matrix according to Eqs.~\eqref{eq_nu1}~--~\eqref{eq_delta} are given by
\begin{align}
\label{twopart_eigensystem}
\nu_1&=\frac{L_A}{L}+\frac{L_A}{L} \sigma, \\ 
\nu_2&=\frac{L_A}{L}-\frac{L_A}{L} \sigma, 
\end{align}
where the overlap $\sigma$ between single particle excitations in the subspace  $A$ is
\begin{equation}
\label{overlap}
\sigma= |\vec{\phi}_1 \vec{\phi}_2|^2=\frac{1}{L_A}\sum_{j \in A} e^{\mathrm{i} (k_1-k_2)j}.
\end{equation}
After the summation of this geometric progression we get: 
\begin{equation}
\label{eq_abs_over}
\sigma=\frac{1}{L_A} \left|\frac{\sin \left[ \frac{L_A (k_1-k_2)}{2}\right]}{\sin \left[ \frac{k_1-k_2}{2}\right]} \right|.
\end{equation}
The EE is then given by: 
\begin{equation}
\label{EE_tb}
S_E^{\text{tb}} =s(\nu_1)+s(\nu_2)
\end{equation}
Setting $L_A/L =1/2$ and $k_1-k_2=n\frac{2 \pi }{L}$, in the thermodynamic limit $L \rightarrow \infty$ we reproduce equations (10) and (11) of the main text. 

In case of arbitrary number of excitations $N$ with momenta $\{k_1,...k_N\}$ the reduced correlation matrix yields: 
\begin{equation}
\label{corr_M_N}
C^A_{ij}= \frac{1}{L} \sum_{m} e^{\mathrm{i} k_m(j-i)} , \quad \text{with}\;i,j \in A
\end{equation}
For its eigenstate $\vec{A}$ we consider the following ansatz: 
\begin{equation}
\label{eigenvector_C}
A_j=\sum_l \alpha_l e^{-\mathrm{i}  k_l j},
\end{equation}
which allows to rewrite the eigenstates equation in the following form: 
\begin{equation}
\label{eigen_Ov}
\mathbf{O} \vec{\alpha} =\nu \vec{\alpha},
\end{equation}
where $\bf{O}$ is a matrix of overlaps with elements: 
\begin{equation}
\label{matrix_overlaps}
O_{ml}=\frac{1}{L_A} \frac{\sin \left[\frac{ L_A (k_m-k_l)}{2} \right]}{\sin \left[\frac{k_m-k_l}{2} \right]}
\end{equation}
So the EE is additive only if all the states are mutually orthogonal within the bipartition $A$. 
\section{Numerical methodology}
\subsection{Numerical data for Fig. 1 of the main part}
The aim is to illustrate the existence of entanglement erasing partner states (EEPS) in Anderson- and many-body localized chains. The joint entanglement entropy (EE) of two simultaneously excited eigenstates depends on their overlap $\vec{\phi}_1 \vec{\phi}_2$ within a bipartition (see main or the above proof). Hence, we conjecture that within localized models, where particles are localized within a localization length of $\xi$ that depends on the employed disorder strength, the erasure of entanglement is (on average) maximized for eigenstates that are localized on spatially adjacent sites. We infer this to be true for $N$ simultaneously excited states: The joint entanglement entropy experiences most erasure if the eigenstates are localized on $N$ adjacent sites in real space. 
Additionally, in order to see the effect, we study eigenstates that are localized as close as possible to the cut between the bipartitions. This is because only such states yield a finite contribution to the EE between both bipartitions. 

Hence, for the non-interacting Anderson chain, we compute all single particle eigenstates $\ket{E_i}$ by means of exact diagonalization. For a given particle number $N$, we search the $N/2$ out of $L$ states that have the largest overlap $\abs{\braket{x_i}{E_j}}^2$ with the $N/2$ lattice sites $\ket{x_i}$ left and right of the cut and compute their Slater determinant. The particle density of this many-particle state is illustrated in the top right inset of Fig.~1 of the main part. The resulting many-particle EE is then readily derived by means of the above employed correlation matrix approach\cite{Peschel2009}.

For the bottom right inset in Fig.~1 of the main text, we study the random field Heisenberg model that exhibits an MBL transition\cite{Luitz2015}. This model corresponds to an interacting fermion model after a Jordan-Wigner transformation. As particle number is conserved, we are again able to search for such eigenstates that have most overlap with the many particle state $$\ket{\psi}=c_{-N/2,}^\dagger c_{-N/2+1}^\dagger \ldots c_{-1}^\dagger c_{0}^\dagger c_{1}^\dagger\ldots c_{N/2-1}^\dagger \ket{},$$
where $c_i^\dagger$ creates a particle on the site $i$ and negative and positive indices correspond to the different bipartitions $\mathcal{H}_A$ and $\mathcal{H}_B$, respectively. This many-particle state is by definition the state with $N$ particles as close as possible to the cut. By studying the properties of the many particle eigenstate $\ket{\phi_i}$ that has most overlap $\abs{\braket{\phi_i}{\psi}}^2$ with $\ket{\psi}$, we expect to see the eigenstate for which the effect of entanglement erasure is maximized. The entanglement entropy of $\ket{\phi_i}$ is then computed regularly by tracing out one bipartition and evaluating the von Neumann entropy of the resulting reduced density matrix.

For both, the interacting and the free model, we employ many disorder ensembles over which we average our results.

\subsection{Numerical data for Fig. 3 of the main part}
In Fig.~3 of the main part, we follow the idea of randomly exciting $N$ single particle states. Again, we solve the Hamiltonians under study by means of exact diagonalization. Then we compare the ratio between the sum of the single particle contributions and the joint entanglement entropy of the corresponding many-particle state. This we perform at various system sizes $L$ and filling ratio $N/L$. This allows us to conduct a finite size scaling and extrapolate our results to the thermodynamic limit. 

\section{Entanglement entropy for the constructed Bell states}
In the main paper we define the two orthonormal bases $\{\ket{a_i}\}$ and $\{\ket{b_i}\}$ of the subspaces $\mathcal{H}_A$ and $\mathcal{H}_B$ that define the bipartition $\mathcal{H} = \mathcal{H}_A\otimes \mathcal{H}_B$. Two Bell-like states 
\begin{align}
\ket{e_{2i}^\text{Bell}} &= \frac{1}{\sqrt{2}} \left( \ket{a_i} + \ket{b_i}\right)\\
\ket{e_{2i+1}^\text{Bell}} &= \frac{1}{\sqrt{2}} \left( \ket{a_i} - \ket{b_i}\right),
\end{align}
yield a maximum entanglement entropy of 1 bit with respect to the above defined bipartition. However, by construction, the total EE vanishes if both states $\ket{e_{2i}^\text{Bell}}$ and $\ket{e_{2i+1}^\text{Bell}}$  are excited simultaneously for a given value of $i$. Thus, they erase each others entanglement entropy, forming entanglement erasing partner states (EEPS). Importantly, two Bell states $\ket{e_{2i}^\text{Bell}}$ and $\ket{e_{2j+1}^\text{Bell}}$ with $i\neq j$ have zero overlap in either part of the bipartition, yielding to zero entanglement erasure.

For a given number of total states $L$, i.e. for $L/2$ Bell-pairs, we now excite $N$ randomly chosen states and ask for the value of the total EE. This value is by construction equal to the number of in how many Bell-pairs $b_i=\{\ket{e_{2i}^\text{Bell}},\ket{e_{2i+1}^\text{Bell}}\}$  \emph{exactly one} of the two states is excited, because only those states give a not-erased contribution to the total EE. We assume $N$ to be an even number. In the case of $s$ single-occupied Bell-pairs $b_i$, there exist $\binom{L/2}{s}$ ways to choose them. Each of such configurations contributes with $2^s$ ways to excite any of the two states of the $s$ pairs. The remaining $N-s$ states will form $(N-s)/2$ double-occupied pairs, which are distributed over the remaining $L/2 - s$ not single-occupied pairs. For this, there are $\binom{L/2-s}{(N-s)/2}$ arrangements. In total, that gives
\begin{equation}
n(s) = 2^s \binom{L/2}{s} \binom{L/2-s}{(N-s)/2}
\end{equation}
possibilities to choose $s$ single-occupied Bell pairs. As the total number of excited states $N$ is even, the number of single occupied Bell states $s$ must be even, too. The expectation value for the number of single occupied states thus yields
\begin{align}
\mean{s} &= \dfrac{ \sum_{s\,\text{even}}^N s\cdot 2^s \binom{L/2}{s} \binom{L/2-s}{(N-s)/2}}{\binom{2L}{N}}\\
&= \frac{(L-N) N}{L - 1},
\end{align}
where the evaluation of the sum is restricted to even values of $s$. Then, the limit 
\begin{equation}
r_\infty^\text{Bell} := \lim_{L\to\infty} \frac{\mean{s}}{N} = \frac{L-N}{L} = 1-(N/L)
\end{equation}
for fixed ratio $N/L$ gives the ratio of entanglement erasure in this model, as presented in Eq.~(17) of the main part.

\end{document}